\documentclass[sn-aps]{sn-jnl}
\usepackage{graphicx}%
\usepackage{multirow}%
\usepackage{amsmath,amssymb,amsfonts}%
\usepackage{amsthm}%
\usepackage{mathrsfs}%
\usepackage[title]{appendix}%
\usepackage{xcolor}%
\usepackage{textcomp}%
\usepackage{manyfoot}%
\usepackage{booktabs}%
\usepackage{algorithm}%
\usepackage{algorithmicx}%
\usepackage{algpseudocode}%
\usepackage{listings}%
\usepackage[normalem]{ulem}
\usepackage{booktabs}  %%%%%%%%%%%
\newcommand\zfig[1]{{\color{blue}#1}}
\usepackage{times}

\begin{document}

\title[]{Mechanical Performance Database for Low-Temperature Alloys}

\author[1]{\fnm{Haoxuan} \sur{Tang}}

\affil[1]{\orgdiv{Applied Mechanics Laboratory, Department of Engineering Mechanics}, \orgname{Tsinghua University}, \city{Beijing}, \postcode{100084}, \country{China}}
%\orgaddress{\street{Street}, 
%\state{State}, 
\author[2]{\fnm{Zhiyuan} \sur{Chen}}
\author[2]{\fnm{Xin} \sur{Yao}}

\affil[2]{\orgname{Beijing Minghui Tianhai Gas Storage and Transportation Equipment Sales Co., Ltd}, \city{Beijing}, \postcode{101100}, \country{China}}

\author*[1]{\fnm{Zhiping} \sur{Xu}}\email{xuzp@tsinghua.edu.cn}

\abstract{
Low-temperature alloys are important for a wide spectrum of modern technologies ranging from liquid hydrogen, and superconductivity to quantum technology.
These applications push the limit of material performance into extreme coldness, often demanding a combination of strength and toughness to address various challenges.
Steel is one of the most widely used materials in cryogenic applications.
With the deployment in aerospace liquid hydrogen storage and the transportation industry, aluminum and titanium alloys are also gaining increasing attention.
Emerging medium-entropy alloys (MEAs) and high-entropy alloys (HEAs) demonstrate excellent low-temperature mechanical performance with a much-expanded space of material design.
A database of low-temperature metallic alloys is collected from the literature and hosted in an open repository.
The workflow of data collection includes automated extraction based on machine learning and natural language processing, supplemented by manual inspection and correction, to enhance data extraction efficiency and database quality.
The product datasets cover key performance parameters including yield strength, tensile strength, elongation at fracture, Charpy impact energy, as well as detailed information on materials such as their types, chemical compositions, processing and testing conditions.
Data statistics are analyzed to elucidate the research and development patterns and clarify the challenges in both scientific exploration and engineering deployment.
}


\maketitle

\clearpage
\newpage

\section*{Background \& Summary}\label{sec1}
\begin{figure*}[h]

{
\includegraphics[width=\textwidth]{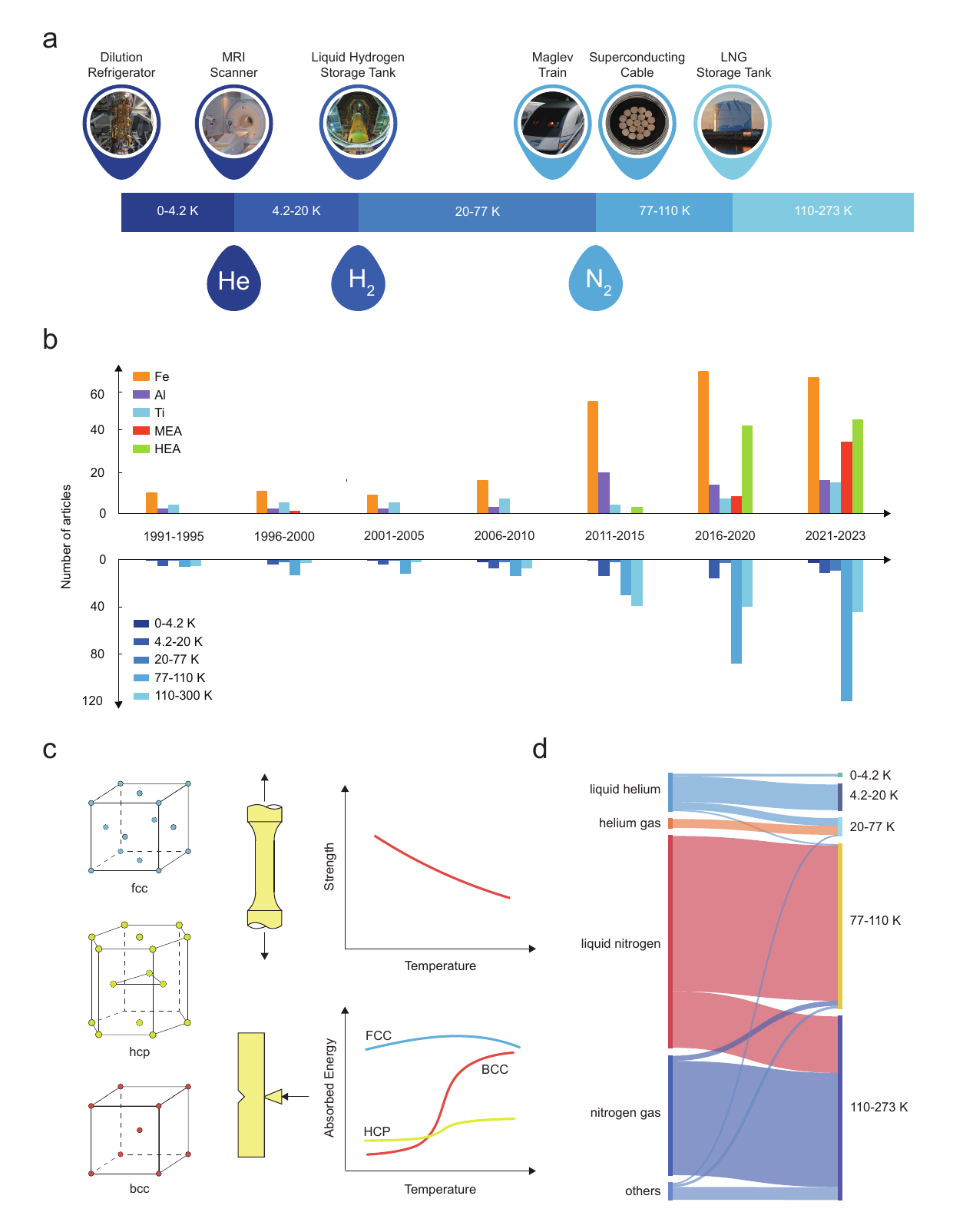}
\caption{
Low-temperature (low-T) technology and applications.
(a) Cryogenic engineering applications, which include a quantum computer (Dmitrmipt, CC BY-SA 4.0 https://creativecommons.org/licenses/by-sa/4.0, via Wikimedia Commons), magnetic resonance imaging (MRI) scanner employing NbTi superconducting magnets (Jan Ainali, CC BY 3.0 https://creativecommons.org/licenses/by/3.0, via Wikimedia Commons), liquid hydrogen storage tank for rockets (NASA Michoud Assembly Facility / NASA/Eric Bordelon, Public domain, via Wikimedia Commons), high-temperature superconducting maglev trains (kallerna, CC BY-SA 4.0 https://creativecommons.org/licenses/by-sa/4.0, via Wikimedia Commons), high-temperature superconducting cables (National Institute of Standards and Technology, Public domain, via Wikimedia Commons), and liquefied natural gas storage tank (Fletcher6, CC BY-SA 3.0 https://creativecommons.org/licenses/by-sa/3.0, via Wikimedia Commons).
The temperature range is not to scale.
(b) Literature data map (1991-2023) ranked by the alloy types and temperatures.
(c) Crystal structures, strength and Charpy impact tests, and representative mechanical properties~\cite{mcclintock1960mechanical}.
(d) Refrigerants and their temperature ranges.
The lengths correspond to the number of data points in the database.
}\label{fig1}
}
\end{figure*}

The term `cryogenics' is defined for applications at $T < 120$ K~\cite{timmerhaus2007cryogenic}.
A wide spectrum of technologies is deployed in low-temperature (low-T) environments, such as quantum technologies including quantum computing and sensing, superconducting-enabled applications such as magnetic resonance imaging (MRI) ($4.2$ K), aerospace or transportation technologies powered by liquid hydrogen (LH2) fuels ($20$ K), and liquefied natural gas storage container ($110$ K) (Fig. \ref{fig1}a).
In these low and discrete temperature ranges, metallic alloys may behave differently from those at room temperature (RT) where material behaviors are better understood.
In practice, achieving precise control of low T poses certain challenges for refrigeration techniques.
In low-T experiments, materials are typically either immersed directly in the refrigerants or cooled by refrigeration systems. 
The latter allows for more accurate temperature control and can achieve even lower temperatures (on the order of milli-Kelvin)~\cite{zu2022development}(Fig. \ref{fig1}d).
OExtreme cold presents significant challenges for maintaining material performance, making material screening and selection critical in the design of low-temperature technologies.
A standardized database with information on material properties and refrigeration techniques can guide the development of low-temperature applications.

For instance, hydrogen energy has recently drawn notable attention for its significance in sustainability development~\cite{yang2022status, zhou2022green}.
LH2 is the main fuel of low-temperature rockets~\cite{cecere2014review}, and nowadays are also widely used in the industry for civil applications~\cite{kumar2023synergy, liu2021production, tollefson2010hydrogen} for their high density and low working pressure~\cite{li2019review}.
In the condition of storage and transportation, the contact of LH2 with metallic containers may result in hydrogen embrittlement and potential risks of catastrophic fracture under loads.
The packaging and reliability design of quantum circuits should also be aware of the material issues at low T~\cite{brecht2017micromachined}.
The mechanical performance of metallic alloys at extremely low-T conditions ($\sim 20$ K) is thus of concern~\cite{qiu2021research}.
Specifically, material selection in engineering is mainly based on the match between the requirement and performance of existing materials that usually depend on the service conditions~\cite{anoop2021review}.
For example, most materials tend to increase their strength and decrease their ductility and toughness at low T~\cite{duthil2015material,mcclintock1960mechanical}.
A ductile-to-brittle transition (DBT) may occur and result in limited yield, which depends on the microstructures of materials~\cite{pineau2016failure}.
This effect is stronger in body-centered cubic (BCC) materials with less ductile performance than that of hexagonal closely-packed (HCP) and face-centered cubic (FCC) materials~\cite{zhang2023unified}(Fig. \ref{fig1}c).

Because of the diversity in the temperature ranges and refrigeration techniques of low-T applications, scientific exploration and technical reports on the mechanical properties of metallic alloys are highly heterogeneous.
Collecting the research data could thus deepen our understanding of material performance for existing applications at cryogenic conditions and elucidate the potential challenges in deploying future technologies.
In this work, we construct a database including tensile and impact test data reported in $715$ scientific articles, which can be used as a reference for the applications of metallic alloys at low-T conditions.

\section*{Methods}\label{sec2}

The workflow follows our previous work on fatigue data~\cite{zhang2023fatigue,zhang2023fatigue2}, which includes content acquisition, data extraction, and database construction.
The database consists of both research metadata and scientific data.
Metadata includes information such as author, publication year, and keywords.
Scientific data includes research data such as material types, chemical compositions, processing and test conditions, and mechanical properties.
For the lack of a unified data description for low-T mechanical properties reported in figures and tables, manual extraction and correction are needed.

\subsection*{Content acquisition}
After collecting the keywords of `low temperature', `alloys', and `mechanical properties', and compiling them into the search formula (Table \ref{table:1}), Web of Science (WoS) returned $8439$ article records and the metadata are obtained through the ‘export’ function.
In search queries, Web of Science automatically applies the `stemming' rules, which reduce word forms to their roots.
The articles are classified through a natural language processing (NLP) classification model~\cite{liu2019roberta} according to their abstracts, the results of which ($715$ articles) are corrected through manual inspection.
The documents are collected through their digital object identifiers (DOIs) in Extensible Markup Language (XML), Hyper Text Markup Language (HTML) or Portable Document Format (PDF) formats.
Articles from Elsevier are downloaded through the official application programming interface (API), while articles from other publishers are automatically downloaded using open-source code \href{https://github.com/olivettigroup/article-downloader}{article-downloader}~\cite{kim2017machine} or manually downloaded from the publishers' websites.
XML/HTML can be automatically parsed and converted into structured text through computer codes, while PDF requires manual processing in large part.

\subsection*{Data extraction}
Images in the articles are directly collected from the XML/HTML files or extracted from the PDF documents using \href{https://pypi.org/project/PyMuPDF/}{PyMuPDF}.
Figures with multiple panels are automatically segmented using a rule-based code.
Figures presenting mechanical properties are screened through a convolutional neural network (CNN) model (ResNet~\cite{he2016deep}).
Strength and fracture elongation data in images are extracted using the MATLAB code \href{https://github.com/xuzpgroup/ZianZhang/tree/main/FatigueData-AM2022/IMEX}{IMageEXtractor}~\cite{zhang2023fatigue2}.
Table data in XML/HTML files are collected by using \href{https://github.com/olivettigroup/table_extractor}{table extractor}~\cite{jensen2019machine} and saved as a JSON format file for subsequent text mining, while table data in PDF files are extracted manually.

Text in the XML/HTML and PDF documents are extracted using \href{https://github.com/xuzpgroup/ZianZhang/tree/main/FatigueData-AM2022/TEXTract}{TEXTract}~\cite{zhang2023fatigue2} and \href{https://github.com/cat-lemonade/PDFDataExtractor}{PDFDataExtractor}~\cite{zhu2022pdfdataextractor}, respectively.
For literature prior to 1991, only scanned image PDFs are available, which were converted into TXT files for further text mining.
Text data are mined using both GPT-$3.5$~\cite{brown2020language} and GLM-$4$ for comparative studies.

The detailed steps for conducting text mining using GPT-$3.5$ are as follows.
The section titles are filtered by keywords (Table \ref{table:2}) to obtain paragraphs related to the experimental methods and data.
GPT-$3.5$ is used to extract information on materials, processing and testing conditions, and mechanical properties from relevant paragraphs or tables.
The prompts in GPT include task descriptions, examples, and the text to be processed.
The task description requires GPT to extract data from the text data and return it in the JavaScript Object Notation (JSON) format.
The examples include paragraphs in the articles and the corresponding uniformly formatted JSON data.
Limited by the maximum tokens in GPT-$3.5$, we provide two examples and process one paragraph or table at a time in data extraction.
Subsequently, the extracted data from each paragraph is manually matched.

In text mining with GLM-$4$, we utilize the same prompts.
GLM-$4$ supports significantly more input and output tokens compared to GPT-$3.5$, allowing the processing of the entire article’s text and tables in a single session.
The textual and tabular data can be automatically aligned, simplifying the subsequent steps of database integration.

The large language model (LLM) approach is challenged by several issues including incorrect or missing contents, heterogeneous data formats, and the lack of capability to extract the composition-processing-testing-performance relationship.
We manually process the data by proofreading the content and formatting the data.
Prompt engineering~\cite{polak2024extracting} and fine-tuning~\cite{min2023recent} techniques should be developed to improve efficiency in the future.

\subsection*{Database integration and data correction}
\begin{figure*}[h]
{
\includegraphics[width=\textwidth]{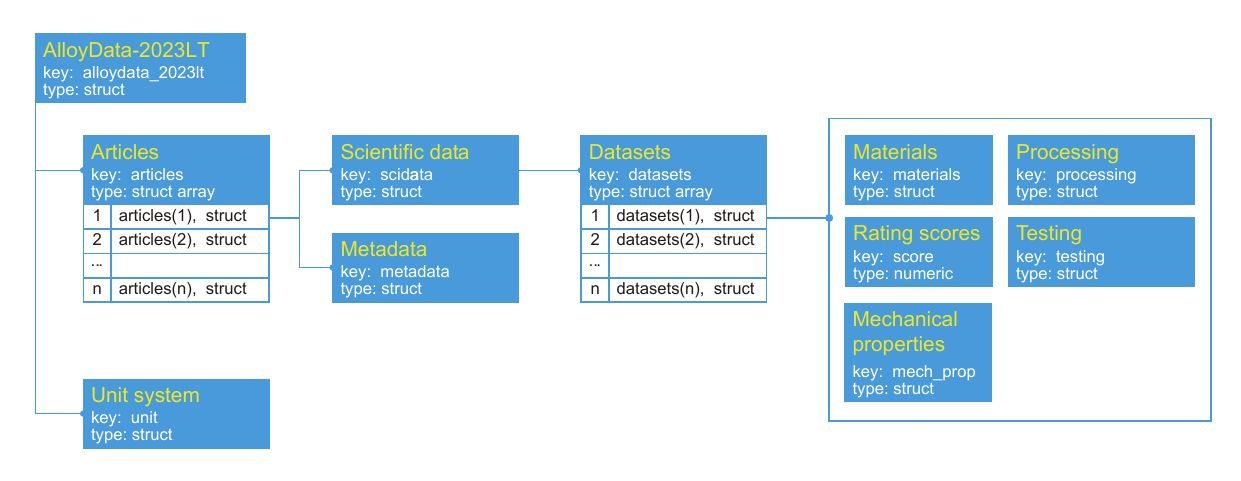}
\caption{
Database structures.
The database of low-T alloys is formatted into a hierarchical tree structure.
The name of each tree node is highlighted in the yellow color.
Keys are defined for easy access by scripts.
Each node has its specific data type.
}\label{fig2}
}
\end{figure*}

To construct the database, mechanical performance data extracted from figures and tables are associated with the data entries extracted from text.
The data is stored in different entries according to the type and composition of materials, processing, and test conditions (\emph{e.g.}, temperatures, sample geometries, refrigeration techniques).
To ensure data quality, manual inspection and correction are carried out for data collected by GPT-$3.5$ and GLM-$4$.
A unified language for `low-T alloys' (ULLTA) is proposed to standardize the data representation (\zfig{Fig. \ref{fig2}}).
The product datasets are exported to a JSON file and proofread by comparing it to the PDF file.

\section*{Data Records}\label{sec3}
Our database compiles literature data from the Web of Science Core Collection up to May 31, 2024 (Fig. \ref{fig1}b).
Compared to materials handbooks~\cite{mcclintock1960mechanical,koshelev1971mechanical} and commercial databases, our database offers several key advantages: it is freely accessible, adheres to the FAIR principles for data sharing, and records full data from the literature, including metadata, materials, processing conditions, and testing conditions - the details often omitted in materials handbooks and commercial databases. 
Unlike static materials handbooks, our database includes the latest research developments, such as high-entropy alloys, and is dynamic and updatable. 
Additionally, the data in materials handbooks is not comprised of original data points but rather conservative fitted curves, limiting its application in engineering references and data-centric research. In contrast, our dataset comprises original experimental data points, along with processing and testing conditions, allowing for better statistical analysis of the effects of composition, processing, and testing conditions on material properties.

\begin{figure*}[h]
{
\includegraphics[width=\textwidth]{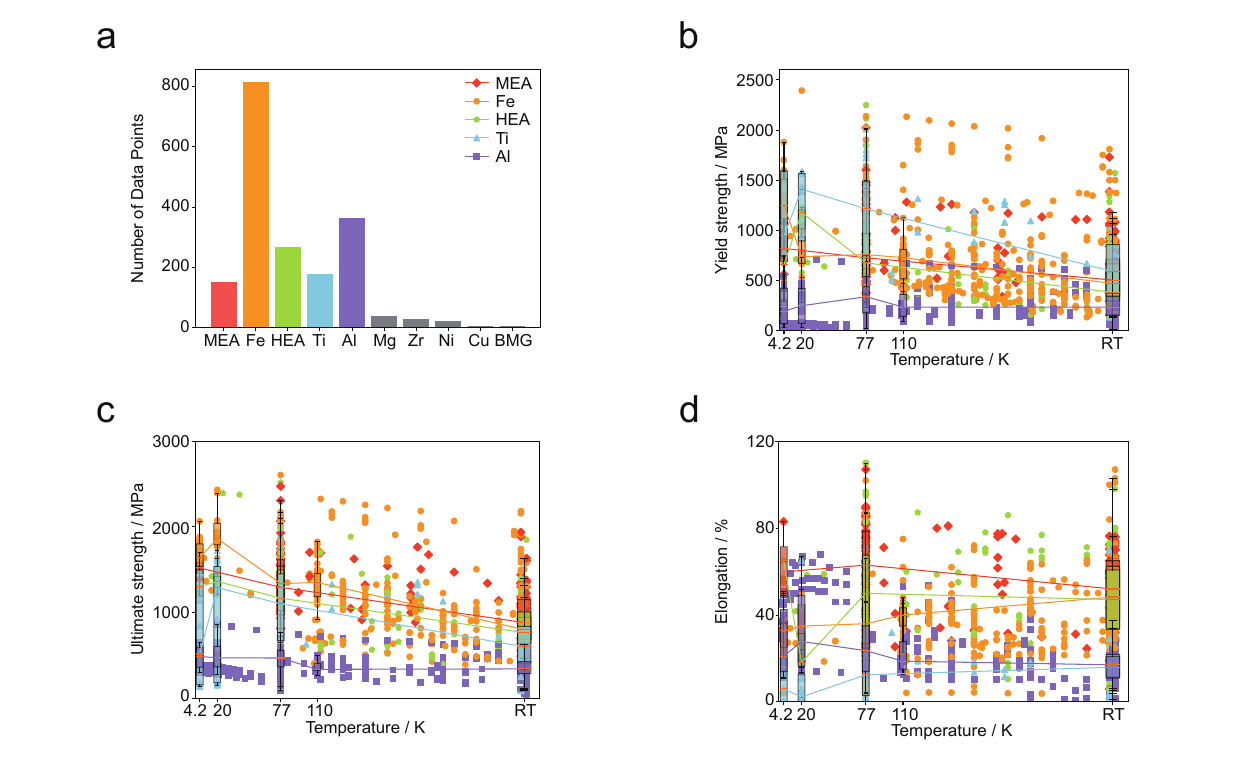}
\caption{
Temperature dependence of mechanical properties of low-T alloys.
(a) Histogram of the number of data points for each alloy.
Representative (b) yield strength-temperature ($\sigma_{\rm y}$-$T$), (c) ultimate tensile strength ($\sigma_{\rm u}$-$T$), and (d) elongation-temperature ($e$-$T$) data of low-T alloys.
Box plots are added at representative temperatures ($4.2$ K, $20$ K, $77$ K, $110$ K, RT).
}\label{fig3}
}
\end{figure*}

\begin{figure*}[h]
{
\includegraphics[width=\textwidth]{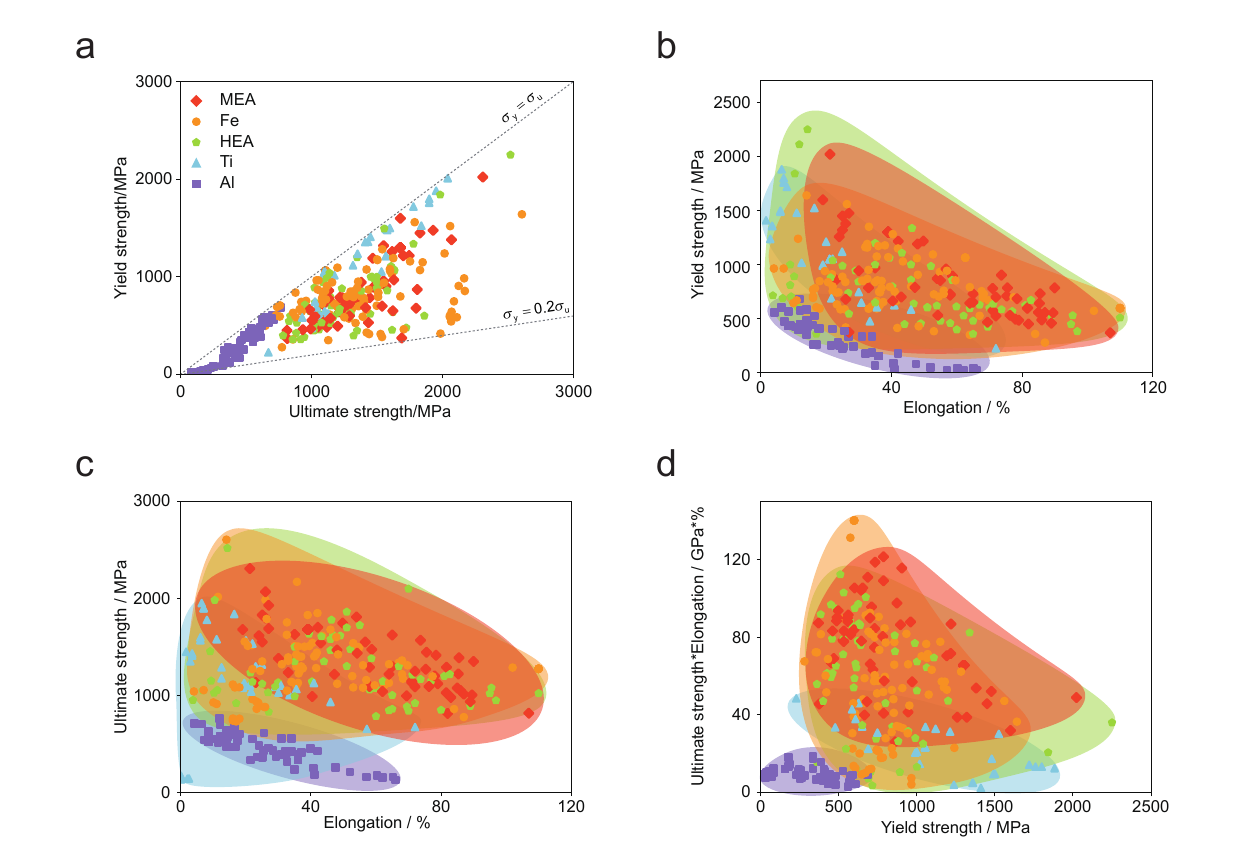}
\caption{
Mechanical properties of low-T alloys at $77$K.
(a) Relation between yield strength $\sigma_{\rm y}$, and UTS, $\sigma_{\rm u}$ at $77$ K. References $\sigma_{\rm y} = \sigma_{\rm u}$ and $\sigma_{\rm y} = 0.2\sigma_{\rm u}$ are added as the dashed lines.
(b-d) Ashby maps in terms of (b) the yield strength versus fracture elongation, (c) the ultimate strength versus fracture elongation, and (d) the product of ultimate strength and fracture elongation versus the yield strength at $77$ K.
Strength-ductility and strength-toughness relationships at $77$ K.
}\label{fig4}
}
\end{figure*}

The statistics of experimental data are summarized in \zfig{Fig. \ref{fig3}}.
The most explored alloys are the MEAs, steels, HEAs, titanium, and aluminum alloys, which are sorted according to the strength-toughness data (Fig. \ref{fig4}d).
The remaining alloys are ranked by the number of reported data points.
From the data contents, we find that steels with FCC structures, such as $300$-series austenitic stainless steels (ASSs) containing Cr and Ni, are the most preferred material for low-T applications due to their excellent low-temperature performance and mature production technology~\cite{qiu2021research}.
Al alloys feature the same FCC structures as ASSs and display no significant DBT.
It also has the advantage of low density and has been widely used in aerospace engineering.
Ti alloys have high specific strengths, good corrosion resistance, and small coefficients of thermal expansion, and are relatively new in aerospace deployment.
$\alpha$- and $\beta$-Ti are HCP and BCC, respectively.
Most titanium alloys used in low-T applications are $\alpha$-phase titanium alloys and two-phase titanium alloys containing a small amount of $\beta$-phase~\cite{zang2022cryogenic}.
One of the latest developments of low-T alloys is the multi-principal element alloys (MPEAs) including medium-entropy alloys (MEAs) and high-entropy alloys (HEAs)~\cite{zhang2021superior,bian2020novel,tong2019outstanding,liu2022exceptional}.
Composition engineering of MPEAs leads to excellent strength-toughness combinations for low-T applications.
At low temperatures, specific compositions of MEA/HEA, such as CoCrFeNi with an FCC structure, exhibit nanotwinning in the later stages of deformation.
The deformation mechanism dominated by twinning imparts ultra-high plasticity to these alloys.
Additionally, the numerous low-energy interfaces generated by the twinning process effectively hinder dislocation movement, increase the work-hardening rate, and delay the onset of necking~\cite{otto2013influences}.
These developments and findings demonstrate the research activity of low-temperature alloys~\cite{wang2024shearing}.

The relationships between the yield strength, tensile strength, fracture elongation, and temperature are plotted for MEAs, steels, HEAs, Ti and Al alloys (Fig. \ref{fig3}b-d).
The mechanical properties are compared at temperatures from $4.2$ K, $20$ K, $77$ K, $110$ K, to RT.
Ti alloys show the highest yield strengths at low T, followed by MEAs, HEAs, and steels with similar yield strengths.
The strengths of Al alloys are the lowest (Fig. \ref{fig3}b).
MEAs, HEAs, steels, and Ti alloys show similar tensile strengths at low T, followed by Al alloys (Fig. \ref{fig3}c).
As the temperature decreases, the overall tensile strengths increase in general.
MEAs and HEAs demonstrate exceptional low-T ductility, with fracture elongation ($\sim60\%$) much higher than that of steels ($\sim40\%$).
Ti alloys exhibit the poorest ductility at low T due to their HCP structures (Fig. \ref{fig3}d).

The low-T data are relatively rich above $77$ K but rare in the ranges of $0-4.2$ K, $4.2-20$ K, and $20-77$ K.
The reason is that liquid nitrogen refrigeration ($> 77$ K) is much more cost-effective than the liquid helium technology ($4.2-77$ K).
There are very few experiments (12/1253) using LH2 refrigeration where the effect of hydrogen embrittlement can be explored, mainly due to safety concerns.
Interestingly, there is a gap between $20$ and $77$ K, where the physics behind the mechanical performance of low-T alloys remains unclear.

Data analysis is first conducted for the data points at $77$ K \zfig{(Fig. \ref{fig4})}.
The ratio $R$ between the yield and tensile strengths is calculated to measure the strength reserve of materials.
As the yield strength approaches the tensile strength, there is either a very short or no yield, and brittle fracture is expected.
A low value of $R$ suggests low strength utilization.
The $R$ values of Al and Ti alloys are close to $1$, indicating their brittleness.
Data dispersion for MPEAs and steels is diverse but suggests better plasticity at low T.
Data in Fig. \ref{fig4} shows that MPEAs and steels exhibit a significant overlap in both the strength-ductility and strength-toughness diagrams, but MPEAs are superior to steels in balancing the trade-offs.

\begin{figure*}[h]
{
\includegraphics[width=\textwidth]{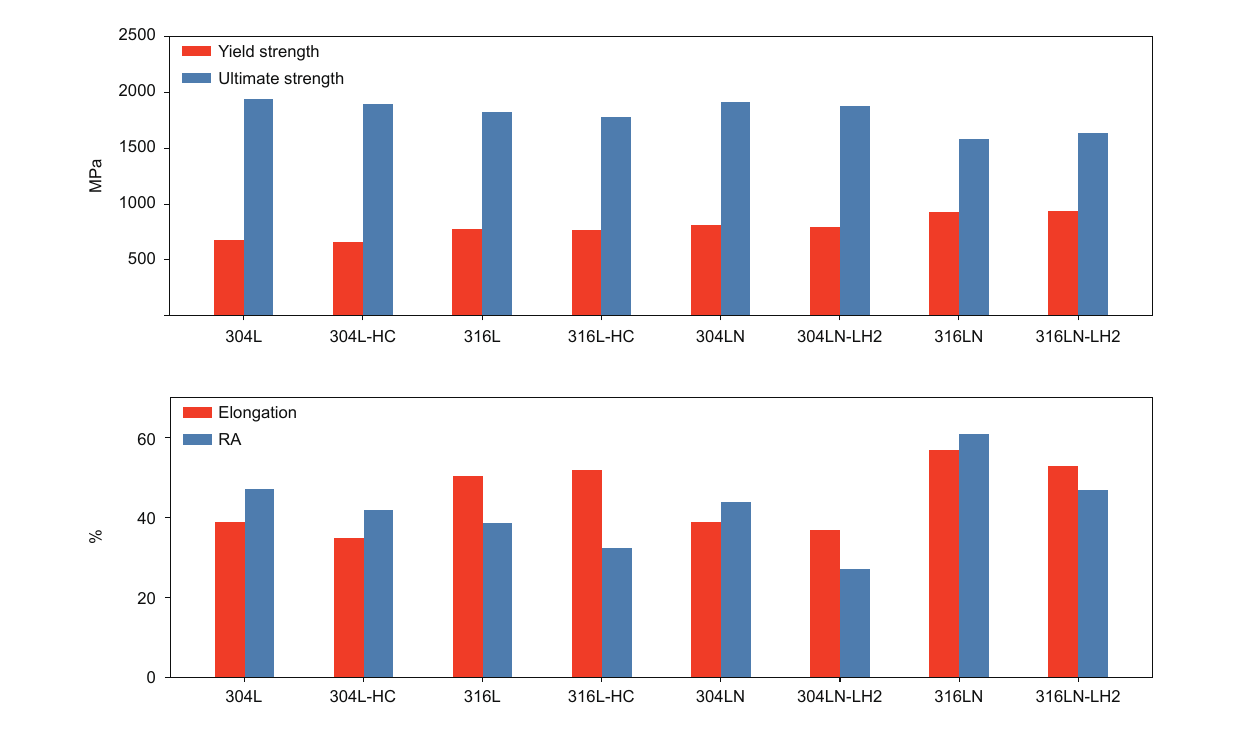}
\caption{
Hydrogen effects on the mechanical properties of low-T alloys~\cite{XU2023131}.
%(a) Mechanisms of hydrogen embrittlement such as hydrogen-enhanced decohesion (HEDE), hydrogen-enhanced localized plasticity (HELP), hydrogen-induced phase transformation, and hydrogen-enhanced strain-induced vacancy~\cite{XU2023131}.
Mechanical properties of selected $300$-series austenitic stainless steels (ASSs) at $20$ K in helium gas, liquid hydrogen (LH2), and hydrogen-charged (HC) conditions.
}\label{fig5}
}
\end{figure*}

The material performance at the LH2 condition is chosen here as a specific example for discussion.
The world is heading for hydrogen, and a large-scale hydrogen economy is essential for a clean energy future~\cite{aziz2021liquid}.
LH2 storage and transportation are promising solutions for large-scale and long-distance applications.
The mechanical performance of $300$-series ASSs under LH2/hydrogen charging (HC) conditions at $20$ K is compared with that in helium conditions \zfig{(Fig. \ref{fig5})}.
The results show that the effect of LH2 and hydrogen filling conditions (\emph{e.g.}, electrochemical hydrogen charging) on the strength and fracture elongation is not significant.
However, the reduction of area (RA) decreases significantly compared with the helium cooling condition, signaling the hydrogen embrittlement effect~\cite{deimel2008austenitic}.
Another study shows that the RA at $80-100$ K hydrogen remains the same as that in a liquid nitrogen environment, and thus no hydrogen embrittlement~\cite{fukuyama2003effect}.
Possible reasons for the observed contradiction include a potential new hydrogen embrittlement mechanism occurring between $20$-$80$ K and the varying susceptibility of alloys to hydrogen embrittlement in different environments, such as gaseous hydrogen, LH2, and electrochemical hydrogen charging.
It should be mentioned that these figures are only for illustration of this \emph{Data Descriptor}, and data analysis can be made through the dataset supplemented with further efforts.

\section*{Technical Validation}\label{sec4}
The performance metrics of figure, text, and table processing show that the F1 scores of automated extraction are $55-92\%$ (Table \ref{table:3}).
The precision of image classification is low ($39\%$).
Therefore, in the process of extracting image data, the image classification results are checked manually to ensure the reliability of the outcomes.
This low precision stems from the lack of standardized image formats for mechanical properties of low-T alloys, which leads to variations in image contents and features.
The use of IMageEXtractor allows automated processing and assisted calibration of the coordinate axes and data points.
To ensure the complete accuracy of the database, the data records extracted from figures, tables, and texts are manually checked and corrected by two individuals in two rounds. 
The accuracy after the first round of verification reached $97\%$, and after the second round, $50$ randomly selected documents are checked, which confirms that the accuracy of extraction is improved to be $100\%$.
%The data records generated from the figures, tables, and texts are then manually examined and corrected to ensure quality.
%$50$ randomly selected documents are checked, which confirms that the accuracy of extraction is improved to be $100\%$.

Incomplete data reporting due to lack of specifications is common in the literature, which brings difficulties to data mining and reduces the credibility or reusability of data.
The mechanical properties commonly reported include temperature, yield strength, tensile strength, and fracture elongation, but some articles may only report part of them.
To evaluate the data completeness, rating scores are calculated as weighted sums of the non-empty entries for each data, which is recorded in the file to facilitate subsequent literature reporting specification and data supplement.

\section*{Usage Notes}\label{sec5}
The AlloyData-2024LT database is available as a JSON file at \href{https://figshare.com/articles/dataset/Low-temperature_Alloy_Mechanical_Properties_Database/25912267}{https://figshare.com/articles/dataset/Low-temperature\_Alloy\_Mechanical\_Properties\_Database/25912267}, which lays the foundation for the material selection and exploration of design spaces of MPEAs, for instance.
The JSON file is formatted into a hierarchical tree structure \zfig{(Fig. \ref{fig2}, Tables \ref{table:3}, \ref{table:4}, and \ref{table:5})}.
The root node is the database, which contains child nodes of articles and the default unit system.
Articles are stored in an array of structures.
Each article contains metadata and scientific data.
Each data set is obtained from experimental tests under different conditions.
The tree node where the data value is stored is called the data entry.
Data entries include string and numeric data types.
Text data is stored as a string.
Data with multiple strings is stored as an array of strings.
The release years of publications are defined as numbers, and other numeric data such as mechanical properties, processing parameters, and experimental temperature are stored in the form of a numeric array.
The tree nodes used to group data entries are called data structures.
A plurality of structures such as articles or data sets are arranged into an array of structures.
To facilitate programming implementation and data acquisition, keys are defined for data entries, structures, and structure arrays.

Users can use the script (ext\_property.m) to access and analyze material data at specific temperatures.
We provide a template script (add\_entry.py) to add entries that are rarely reported such as the grain sizes and moduli.
Our database's data can be formatted as ULLTA and directly imported using the script (import\_ullta.py).
For dataset scoring, we offer a script (cal\_rate\_score.py) for users to customize weights.

\backmatter

\section*{Code availability}
Extraction of information from texts, figures, and tables is based on open-source codes and models such as \href{https://simpletransformers.ai/}{Simple Transformers}, ResNet, and \href{https://github.com/olivettigroup/table_extractor}{table extractor}~\cite{he2016deep}.
In-house scripts for data extraction are publicly released with our previous work~\cite{zhang2023fatigue,zhang2023fatigue2} and at \href{https://github.com/xuzpgroup/HaoxuanTang/tree/main/LowTData}{https://github.com/xuzpgroup/HaoxuanTang/tree/main/LowTData}), which can be used by acknowledging the current article and under the CC-BY license.

\section*{Acknowledgements}
The work was supported by Beijing Municipal Science and Technology Commission via grant Z231100007123015 and National Natural Science Foundation of China via grants 12425201 and 52090032.

\section*{Author information}
\subsection*{Contributions}
Z.X. conceived and supervised the research.
H.T. performed the work.
All authors participated in discussing the results and preparing the manuscript.

\section*{Ethics declarations}
\subsection*{Competing interests}
The authors declare no competing interests.

\clearpage
\newpage
\begin{table}[h]
\centering
\begin{tabular}{|l|l|}
\hline
Category & Keyword  \\
\hline
temperature & cryogenic temperature/low temperature/cryogenic environment  \\
\hline
property & mechanical behavior/mechanical property  \\
\hline
alloy & alloy/steel/aluminum/titanium  \\
\hline
\end{tabular}
\caption{Keywords used for article search in the citation database.}
\label{table:1}
\end{table}

\clearpage
\newpage
\begin{table}[h]
\centering
\begin{tabular}{|l|}
\hline
Keyword \\
\hline
test/experiment/method/material/process/manufacture/character/fabrication/property/mechanic/specimen/sample \\
\hline
\end{tabular}
\caption{Keywords used for paragraph screening.}
\label{table:2}
\end{table}

\clearpage
\newpage
\begin{table}[h]
\centering
\begin{tabular}{|l|l|l|l|l|}
\hline
Source & Function & Precision & Recall &F1  \\

\hline
\multirow{2}{*}{figure} & figure segmentation& 89 & 94& 92 \\
\cline{2-5}
& figure classification& 39 &95 & 55 \\
\hline
GPT-$3.5$& data extraction&  82&97 & 89 \\
\hline
GLM-$4$& data extraction&  85&97 & 91 \\
\hline
\end{tabular}
\caption{Evaluation metrics of automated data processing.}
\label{table:3}
\end{table}

\clearpage
\newpage
\begin{table}[h]
\centering
\begin{tabular}{|l|l|l|l|}
\hline
Struct & Data Entry & Data Key & Data Type  \\
\hline
\multicolumn{4}{|l|}{Metadata}  \\
\hline
\multirow{5}{*}{} &Title&title&string \\
\cline{2-4}
& Authors&author&string  \\
\cline{2-4}
& Source of the publication&source&string  \\
\cline{2-4}
& Year of publication&year&numeric  \\
\cline{2-4}
& DOI&doi&string  \\
\hline
\end{tabular}
\caption{Contents of the ‘metadata’ struct.}
\label{table:4}
\end{table}

\clearpage
\newpage

\begin{table}[h]
\centering
\begin{tabular}{|l|l|l|l|}
\hline
Struct & Data Entry/Struct & Data Key & Data Type  \\
\hline
\multicolumn{4}{|l|}{Materials}  \\
\hline
\multirow{4}{*}{} & Type of the materials & mat\_type & string \\
\cline{2-4}
& Name of the materials&mat\_name&string  \\
\cline{2-4}
& Composition&composition&numeric  \\
\cline{2-4}
& Type of the element ratio of composition&ratio\_type&string  \\
\hline
\multicolumn{4}{|l|}{Processing}  \\
\hline
\multirow{4}{*}{} &Processing parameters&proc\_para&sturct array \\
\cline{2-4}
& Surface treatment parameters&surf\_para&struct array  \\
\cline{2-4}
& Ingot description&ingot\_desc&string  \\
\cline{2-4}
& Size of ingot&ingot\_size&numeric \\
\hline
\multicolumn{4}{|l|}{Testing}  \\
\hline
\multirow{4}{*}{} &Types of tests&test\_type&string \\
\cline{2-4}
& Test temperature&test\_tem&numeric  \\
\cline{2-4}
& Test environment&test\_env&string  \\
\cline{2-4}
& Refrigerant&refrigerant&string \\
\cline{2-4}
& Test machine&test\_mac&string  \\
\cline{2-4}
& Test standard&test\_standard&string  \\
\cline{2-4}
& Load control&load\_ctrl&string  \\
\cline{2-4}
& Rate&rate&numeric  \\
\cline{2-4}
& Specimen description&spec\_desc&string \\
\cline{2-4}
& Cross-section shape&spec\_shape&string \\
\cline{2-4}
& Cross-section size of specimens&spec\_size&numeric \\
\cline{2-4}
&Specimen standard&spec\_standard&string \\
\cline{2-4}
&Specimen direction&spec\_dir&string \\
\cline{2-4}
&Specimen notch&spec\_notch&string \\
\hline
\multicolumn{4}{|l|}{Mechanical properties}  \\
\hline
\multirow{4}{*}{} &Temperature&temperature&numeric \\
\cline{2-4}
& Yield strength&yield\_strength&numeric  \\
\cline{2-4}
& Ultimate strength&ultimate\_strength&numeric  \\
\cline{2-4}
& Fracture elongation&elongation&numeric \\
\cline{2-4}
& Impact energy&impact\_energy&numeric \\
\hline
\end{tabular}
\caption{Contents of the ‘datasets’ struct array.}
\label{table:5}
\end{table}

\clearpage
\newpage

\clearpage
\newpage

\clearpage
\newpage

\clearpage
\newpage

\clearpage
\newpage

\clearpage
\newpage

\bibliography{main_text}

\end{document}